\documentclass[sort&compress,english,preview,5p]{elsarticle}

\usepackage{lineno,hyperref}
\modulolinenumbers[5]

\journal{Ultramicroscopy}

\usepackage{graphicx}
\usepackage{xcolor}
\usepackage{tabularx}
\usepackage{amsmath}
\usepackage{physics}
\usepackage{float}

\begin{document}

\begin{frontmatter}

\title{Interactive event-driven 4D-STEM with Timepix4 and live analytical ptychography}

\author[emat,nanolab]{Arno Annys\corref{cor1}}
\author[emat,nanolab]{Hoelen L. Lalandec Robert}
\author[emat,nanolab]{Jo Verbeeck}

\cortext[cor1]{Corresponding author. Email: Arno.Annys@uantwerpen.be}
\address[emat]{Electron Microscopy for Materials Science (EMAT), University of Antwerp, Groenenborgerlaan 171, 2020 Antwerp, Belgium}
\address[nanolab]{NANOlight Center of Excellence, University of Antwerp, Groenenborgerlaan 171, 2020 Antwerp, Belgium}

\begin{abstract}
Four-dimensional scanning transmission electron microscopy (4D-STEM) provides access to rich diffraction information at every probe position, but the high data rates associated with pixelated detectors remain a major challenge for interactive operation. Event-driven detectors offer an efficient alternative by representing the measurement as a sparse stream of individual incidences, substantially reducing requirements related to data handling and enabling on-the-fly processing. Here, we develop a high-throughput, real-time, 4D-STEM workflow based on the Timepix4 detector and demonstrate its operation at microsecond dwell times and standard beam currents. A common external clock shared by the scan unit and detector allows the probe position of each electron event to be determined directly from its time-of-arrival, eliminating the need for inter-stream synchronization and enabling fully concurrent and independent multi-stream processing. The resulting pipeline continuously processes the detector data during acquisition, with the reconstructed images directly available for experimental feedback. We demonstrate live virtual annular detector imaging, Real-Time Integration Center of Mass (RiCOM) and direct electron ptychography, using the Wigner distribution deconvolution (WDD) approach within the guided progressive reconstructive imaging (GPRI) framework. The workflow sustains the high event rates of Timepix4 while maintaining zero latency for live interactive operation, thus extending event-driven 4D-STEM towards beam currents and acquisition conditions beyond those previously demonstrated with Timepix3-based systems. These results demonstrate that the combination of high-bandwidth event-driven detection, hardware-level synchronization and parallel processing can transform 4D-STEM from an offline technique into an interactive imaging modality, in which dose-efficient phase contrast and diffraction-based signals can serve directly as immediate feedback during STEM experiments.
\end{abstract}

\begin{keyword}
    4D-STEM, Event-driven Detection, Timepix4, Ptychography
\end{keyword}

\end{frontmatter}

\section*{Introduction}
4D-STEM\cite{Yang2015a,Muller2016} records a diffraction pattern at every probe position in scanning transmission electron microscopy (STEM) experiments, providing a rich dataset for applications ranging from nano-beam electron diffraction \cite{Uesugi2011,Beche2009,Wu2023b,Kobler2013,Viladot2013} to high-resolution phase contrast imaging. In particular, electron ptychography \cite{Rodenburg2019,Wang2025} has enabled high-resolution phase imaging with strong sensitivity to light elements and high dose efficiency \cite{OLeary2021,Yang2016,Yang2017,Wang2017a,Leidl2023,LalandecRobert2025}, while providing access to spatial frequencies beyond conventional imaging limits
\cite{Maiden2011,Nellist1995,Jiang2018}. Analytical approaches such as Wigner distribution deconvolution (WDD) \cite{Rodenburg1992,McCallum1992,Nellist1994,Li2014} and sideband integration (SBI) \cite{Rodenburg1993,Pennycook2015,Yang2015b,Yang2016a,LalandecRobert2025} provide computationally efficient routes to reconstruct phase contrast directly from 4D-STEM measurements, making them particularly attractive for live imaging.
\\
\\
The rapid development of 4D-STEM is closely linked to advances in direct electron detectors (DED) \cite{Levin2021,McMullan2007}. Hybrid pixel array detectors (HPAD), in particular, have enabled high-speed acquisition of diffraction patterns with high detective efficiency and frame rates exceeding 100~kHz \cite{Zambon2023}. In a conventional frame-based acquisition, however, the data volume per frame is fixed by the detector dimensions, irrespective of the number of electrons detected. At high frame rates, or equivalently short dwell times, this results in a rapidly increasing data rate and an inefficient representation of the sparse electron signal.
Event-driven detectors, such as those based on Timepix3 \cite{Poikela2014,Frojdh2015} and Timepix4 \cite{Llopart2022}, offer a fundamentally different approach. Rather than recording complete frames, each detected electron generates an event containing its pixel coordinates, time-of-arrival (TOA) and time-over-threshold (TOT). The resulting sparse representation scales with the number of detected electrons rather than with the detector size and acquisition rate, while enabling continuous acquisition. Hence, the volume of data generated, with its associated memory, bandwidth, and processing requirements, are substantially reduced. These properties make event-driven detection particularly attractive for high-speed and low-dose 4D-STEM, and have enabled live processing workflows that would be difficult to realize using conventional frame-based representations \cite{Jannis2022,ANNYS2025}.
\\
\\
Despite the increasing ability of 4D-STEM methods to extract information from rich diffraction datasets, the processing is still commonly performed separately from the experimental session. As a consequence, even when the 4D-STEM dataset contains the information ultimately sought by the experimentalist, a conventional single-pixel signal such as high-angle annular dark-field (HAADF) often remains the primary signal used to operate the microscope and navigate the specimen. This limits the feedback available during the experiment: tasks such as locating a region of interest, identifying a zone axis, optimizing aberrations and focusing should ideally be guided by the 4D-STEM information itself. Images reconstructed from the data might reveal different contrast mechanisms, with higher contrast and differing focus- and thickness-dependencies \cite{Robert2021,Gao2022,Gao2024,Li2022a}.
The practice of real-time reconstruction can therefore change not only how the diffraction patterns are processed, but how the technique is used experimentally. Several workflows for live phase contrast imaging from 4D-STEM data have been demonstrated, including iterative ptychography \cite{Weber2024}, SBI \cite{Strauch2021}, WDD \cite{Bangun2023,LALANDECROBERT2026114411}, and iCoM \cite{Yu2022a}. To provide feedback comparable to conventional interactive STEM imaging, however, the reconstruction must keep pace with the acquisition and operate with sufficiently low latency. This places stringent requirements on both the computational efficiency of the reconstruction and the data pathway from detector to image. Event-driven pipelines are particularly well-suited to this task because the sparse representation allows the complete acquisition and reconstruction to be performed directly at the most elementary level. That is, individual electron counts. RiCOM \cite{Yu2022a}, for example, enables processing at microsecond dwell times using a fully sparse pipeline. More recently, the guided progressive reconstructive imaging (GPRI) framework \cite{LALANDECROBERT2026114411} demonstrated fully event-driven real-time analytical ptychography using WDD and SBI. These developments establish the feasibility of live 4D-STEM, but also highlight the importance of matching the processing architecture to the operation mode of the detector.

\section*{Motivation of this work}
Our previous work demonstrated that event-driven acquisition and processing can remove many of the constraints associated with conventional dense 4D-STEM approaches\cite{ANNYS2025,Gholam2026}. By processing the data directly in their native, sparse, event-based representation, the requirements for memory, data transfer, storage, and computation can be substantially reduced, while also enabling real-time processing. In particular, live analytical ptychography was demonstrated using a Timepix3-based detector \cite{LALANDECROBERT2026114411}. However, sparse and continuous operation came at the cost of a different limitation: the maximum electron flux of the Timepix3 constrained the acquisition rate and, consequently, the range of operating conditions under which real-time event-driven 4D-STEM could be performed.
Timepix4 substantially alters this balance. Its increased readout bandwidth allows event-driven acquisition at electron fluxes reaching the range of tens of pA, significantly extending the operating range compared with Timepix3\cite{Llopart2022,SKOUMAL2026114410,Dimova2024}. Recently, some first experimental demonstrations of 4D-STEM using Timepix4 have been reported\cite{SKOUMAL2026114410,kumar2026,Dushimineza2026}. In principle, this creates the possibility of performing 4D-STEM without many of the practical constraints that have traditionally separated pixelated-detector experiments from conventional single-pixel STEM workflows. One could envision fast 4D-STEM experiments in which the diffraction data is acquired continuously, processed directly in the sparse representation and reconstructed sufficiently quickly to provide the same immediate experimental feedback that is routinely available from signals such as HAADF.

The increased bandwidth of modern event-driven detectors does not by itself guarantee an interactive 4D-STEM workflow. As the achievable event rate increases, the acquisition and processing pipeline must be designed to sustain this rate while maintaining sufficiently low latency for experimental feedback. This becomes particularly relevant for detectors that distribute the event data over multiple parallel streams, where the relationship between the streams and the synchronization with the scanning system can impose additional constraints on the processing architecture. Whether these streams can be processed independently, or instead require communication and synchronization between processing threads, is therefore an important consideration in the design of a real-time event-driven 4D-STEM workflow.

Thus, as event-driven detectors move towards higher flux and bandwidth, the design of the complete acquisition and processing workflow becomes increasingly important. A detector capable of providing the required data rate alone does not guarantee a real-time 4D-STEM experience; the acquisition, synchronization, data handling and reconstruction must be designed as a coherent system. In this work, we develop such a workflow for Timepix4-based 4D-STEM. We first introduce a parallel processing architecture for the 32 event streams produced by the detector and investigate the role of synchronization in enabling efficient stream-independent processing. We then demonstrate the resulting workflow experimentally, including live analytical ptychographic reconstruction at the microscope. Together, these results establish the practical framework required to approach an interactive 4D-STEM workflow in which the rich information of the full diffraction pattern can be used during the experiment without imposing the data-handling constraints traditionally associated with 4D-STEM.

\section*{Timepix4 data pipeline}
The Timepix4 chip is commercially available for TEM applications, as part of the Merlin T4 DED\cite{MerlinT4} (Quantum Detectors). In its raw data format, each detected event is represented by a 64-bit data packet. Of these bits, 18 encode the pixel coordinate, 10 encode the time-over-threshold (TOT), 1 indicates pile-up, and 16 encode the coarse time-of-arrival (TOA) with a temporal resolution of 25~ns. The remaining bits contain additional information for refined time stamping. Auxiliary packets extend the TOA range from approximately 1.6~$\mu$s to several days. Timepix4 provides a seamless $448 \times 512$ pixel detection area. The readout electronics separates two chip halves, each covering one $448 \times 256$ half of the detector. Each half distributes its data over 16 digital event streams, resulting in a total of 32 independent event streams. This differs substantially from typical Timepix3-based workflows, where event data are commonly provided by the control computer as a single stream. The 32 parallel streams are a deliberate architectural design feature enabling large readout bandwidth, which the processing software leverages for parallelization. This requires consideration in the design of the data processing pipeline, as the relationship between the individual streams must be considered when assigning events to their corresponding scan positions.
\\
The event streams are continuously written to HDF5 files on a RAM disk. This approach allows newly acquired data to be accessed while the measurement is still in progress, all the while retaining the complete dataset for subsequent storage, if needed. More importantly, the parallel nature of the detector output provides an opportunity to process the 32 streams concurrently. Specifically, if the streams can be processed independently, each can be assigned to a separate processing thread, thus allowing the processing throughput to scale with the available computational resources. If, however, information must be exchanged between streams, the parallelization becomes more complex due to the need for thread synchronization and the potential for data races.

Consequently, the synchronization strategy between the scan unit and detector is not only relevant for the accurate assignment of events to probe positions, but also directly determines the degree to which the 32 event streams can be processed independently. The different synchronization strategies and their implications for the real-time processing architecture are discussed in the following section.

\section*{Synchronization in event-driven 4D-STEM}
The TOA of each detected electron can be used to determine its corresponding real-space probe position $(r_x,r_y)$, provided that the detector time base is related to the timing of the scan system. The synchronization strategy used to establish this relation has a direct impact not only on the accuracy of the probe position assignment, but also on the design and parallelization of the real-time processing pipeline. In the following, we consider three cases: an unsynchronized detector and scan system, synchronization using triggers, and synchronization using a common external clock.
\subsection*{Asynchronous case}
\begin{figure}[h]
    \centering
    \includegraphics[width=\linewidth]{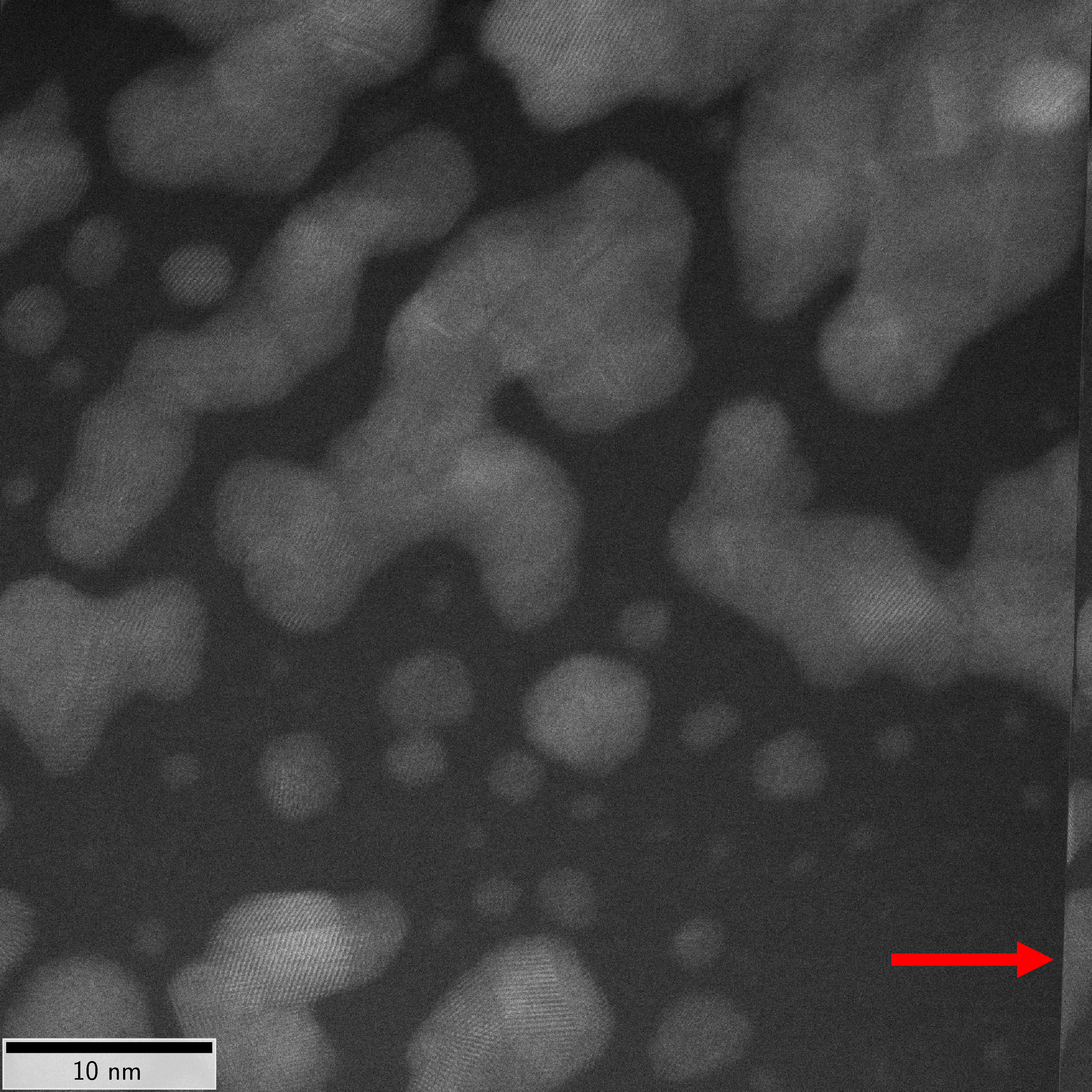}
    \caption{Experimental demonstration of image distortion caused by clock deviation between the detector and the scan unit in the absence of synchronization. The specimen consists of gold nanoparticles on a standard calibration grid. The scan consisted of 2048 $\times$ 2048 scan points using a pixel dwell time of 10~$\mu$s. The clock deviation in this image is estimated to be 14 ppm.}
    \label{fig:fig_clock_drift}
\end{figure}

It may be tempting to omit explicit synchronization between the detector and the scan engine, particularly for fast measurements. However, the independent crystal oscillators used by the two systems are subject to temperature-dependent frequency variations and can exhibit frequency deviations on the order of parts per million (ppm) \cite{Walls1995}. Even when the total acquisition time is short, this frequency difference accumulates into a measurable timing error. Without synchronization, the TOA measured by the detector therefore gradually deviates from the time base of the scan engine. When the probe position is inferred from the detector TOA, this deviation translates directly into an error in the assigned probe position. The resulting positional error accumulates throughout the scan, leading to a progressive distortion of the reconstructed image as demonstrated in figure \ref{fig:fig_clock_drift}.
\subsection*{Triggers}
Triggers are commonly used to synchronize the scan system and the detector, for example by generating a trigger at the end of each scan pixel or line. In an event-driven detector, a trigger can be encoded as a unique event in the detector data stream and is therefore timestamped using the detector clock. The trigger events provide reference points between which the relative timing of the detector and scan unit can be determined. As long as the interval between consecutive triggers is sufficiently short, the clock deviation between the two systems can be considered negligible, allowing the probe position of electron events to be determined from their TOA relative to the most recent trigger.

In a real-time processing pipeline, however, the use of triggers introduces an additional synchronization problem. The Merlin T4 produces 32 independent event streams, and a trigger generated by the scan unit can appear in any of the 16 streams associated with the chip half that handles the triggers. Consequently, the processing of an individual stream cannot be performed independently: a processing thread must be aware of trigger events that may be detected by the other threads. This is particularly relevant when the trigger is used to determine a timing correction. Hence, while the electron events themselves can be processed independently, the timing information required to assign them to the correct probe position has to be shared between the streams.

In principle, the TOA information is not required to determine the probe position in a pixel trigger-based acquisition, once the trigger boundaries are known. The triggers define the temporal boundaries of the scan positions. In a conventional offline workflow, this dependency can be handled using a two-pass approach. During a first pass through the data, all trigger events are identified and their timestamps determined. In a second pass, the 32 data streams can then be processed independently, since the locations of all relevant triggers are already known. Although straightforward, this two-pass approach is undesirable for a real-time workflow. It introduces an inherent delay between data acquisition and reconstruction, while requiring the complete dataset to be traversed twice. For the large data volumes generated by Timepix4 detectors, the computational cost of this additional pass can be substantial.

A real-time, single-pass implementation therefore requires the trigger information to be shared between the processing threads. One possible approach is to maintain a thread-safe timing correction factor that represents the difference between the expected and measured TOA of the most recent trigger. Whenever a processing thread encounters a trigger, this correction factor is updated. All processing threads can then use the most recent correction factor when processing subsequent electron events. If triggers are generated sufficiently frequently, for example at the end of every scan line, the clock deviation accumulated between consecutive corrections can be kept sufficiently small. Each electron event can consequently be employed using the current timing correction, and its probe position calculated from the corrected TOA.
This approach effectively implements, as part of the processing software, a continuously updated approximation of a common time base. It allows the individual event streams to be processed concurrently while accounting for the relative clock drift between the detector and scan unit. However, unlike a hardware-dependent common time base approach, the processing threads remain coupled through the shared timing correction, requiring careful consideration of thread synchronization and potential data races. An external clock approach avoids this additional complexity by eliminating the relative clock drift at the hardware level, as discussed in the following section.
\subsection*{External clock}
A straightforward way to ensure that the scan unit and the DED remain synchronized in time, and consequently maintain a consistent association between detection events and probe positions, is to operate both systems from a common clock. Such a synchronization scheme can be implemented using a central clock, which is common practice in particle physics, allowing to synchronize many devices\cite{Serrano2009}. Here, we use an approach previously demonstrated for Timepix3 \cite{Jannis2022}, and adapt it to the Timepix4 detector. In this implementation, the scan unit provides a 40~MHz clock that serves as the external clock for the detector. As a result, the time base used to timestamp the events is directly derived from the clock governing the scan.
This common time base substantially simplifies the processing of the multiple detector data streams. Since the scan unit and detector remain synchronized, the probe position corresponding to each electron event can be determined directly from its time-of-arrival (TOA), without requiring information from the other data streams. 

Importantly, because the probe position can be calculated independently for every event, no synchronization or communication between the processing threads is required to determine the real-space location to which an event belongs. This makes the external clock approach particularly well-suited to the parallel processing of the 32 independent data streams produced by the Merlin T4.

\section*{Processing pipeline}
\begin{figure}[H]
    \centering
    \includegraphics[width=\linewidth]{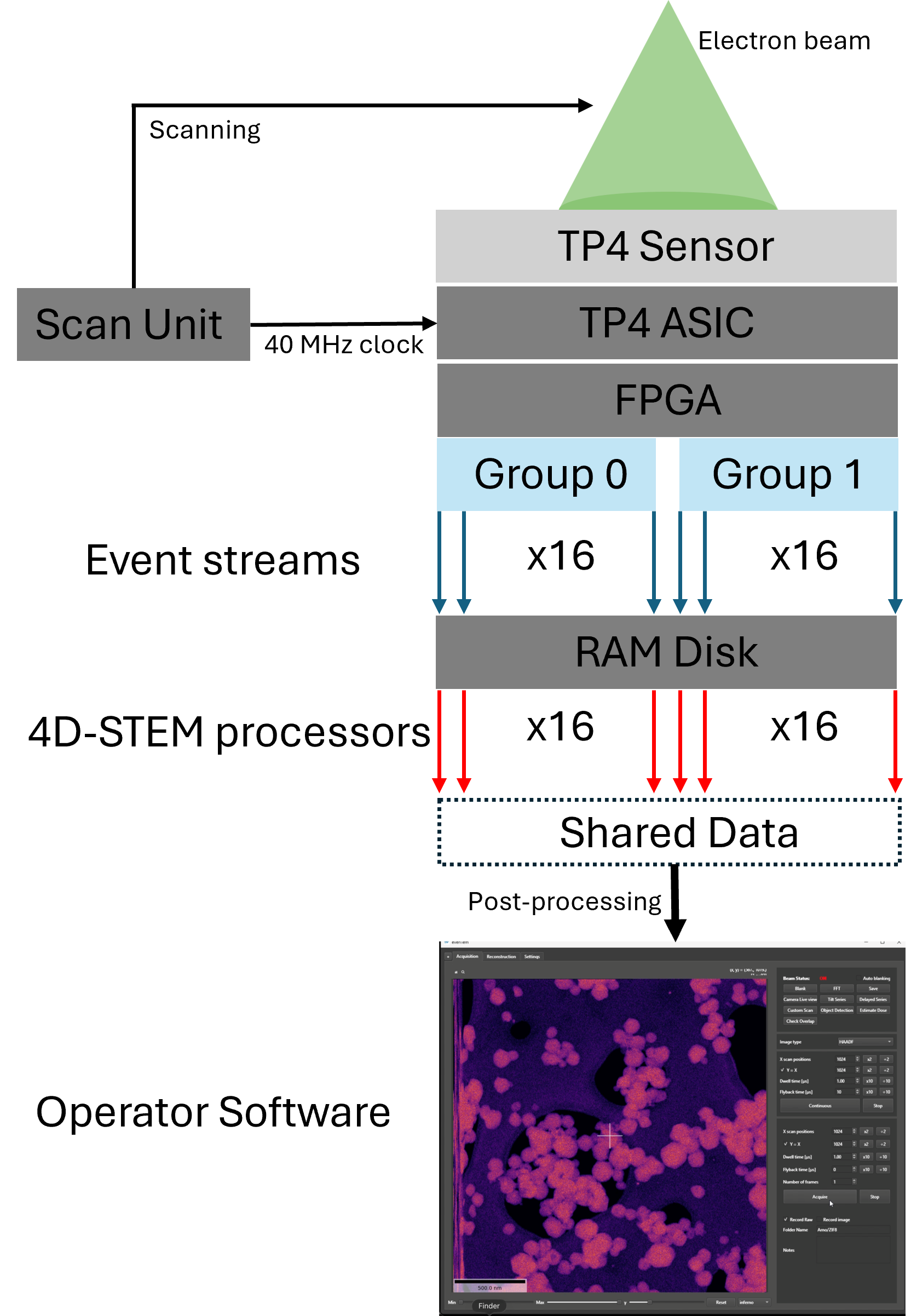}
    \caption{Schematic overview of the external clock workflow. The scan unit controls the beam position and ensures detector consistency by providing the external clock for the Timepix4. The raw event data is written to 32 independent files on the RAM disk. Each file is then associated with its own independent data processor. All 32 data processors have thread-safe access to shared data, such as a reconstructed image, that can be visualized in the operator software.}
    \label{fig:processingpipe}
\end{figure}

The processing of a single stream of events is relatively straightforward and has been described previously for both externally clocked and trigger-based workflows \cite{Jannis2022,ANNYS2025}. For multiple related event streams, however, the design of the processing pipeline is strongly determined by the synchronization strategy, as discussed above. Figure \ref{fig:processingpipe} illustrates the processing pipeline for the 32 data streams generated by the Merlin T4, based on external clock synchronization, which allows all of them to be processed independently. The raw event streams produced by the detector are written to a RAM disk, as continuously growing HDF5 files. Since newly acquired data are appended sequentially to these files, events can be processed while the measurement is still ongoing, without waiting for the acquisition to finish. A dedicated processing thread is thus assigned to each of the 32 data streams, which independently reads and processes the corresponding events as they become available.

The processing threads concurrently update shared data, containing the final measurement output. This shared data is implemented using thread-safe C++ atomic variables, allowing contributions from different streams to be accumulated without introducing data races. This architecture separates the acquisition and processing of the individual detector streams while maintaining a common, continuously updated reconstruction, thus enabling the 4D-STEM data to be processed and visualized in real-time during the experiment.
The workflow is centrally orchestrated through the EvenTem software\cite{ANNYS2025}, which communicates with the vendor APIs for hardware control, ensuring proper synchronization, data management, real-time processing and visualization.
\section*{Results}
\begin{figure*}[h]
    \centering
    \includegraphics[width=1.0\textwidth]{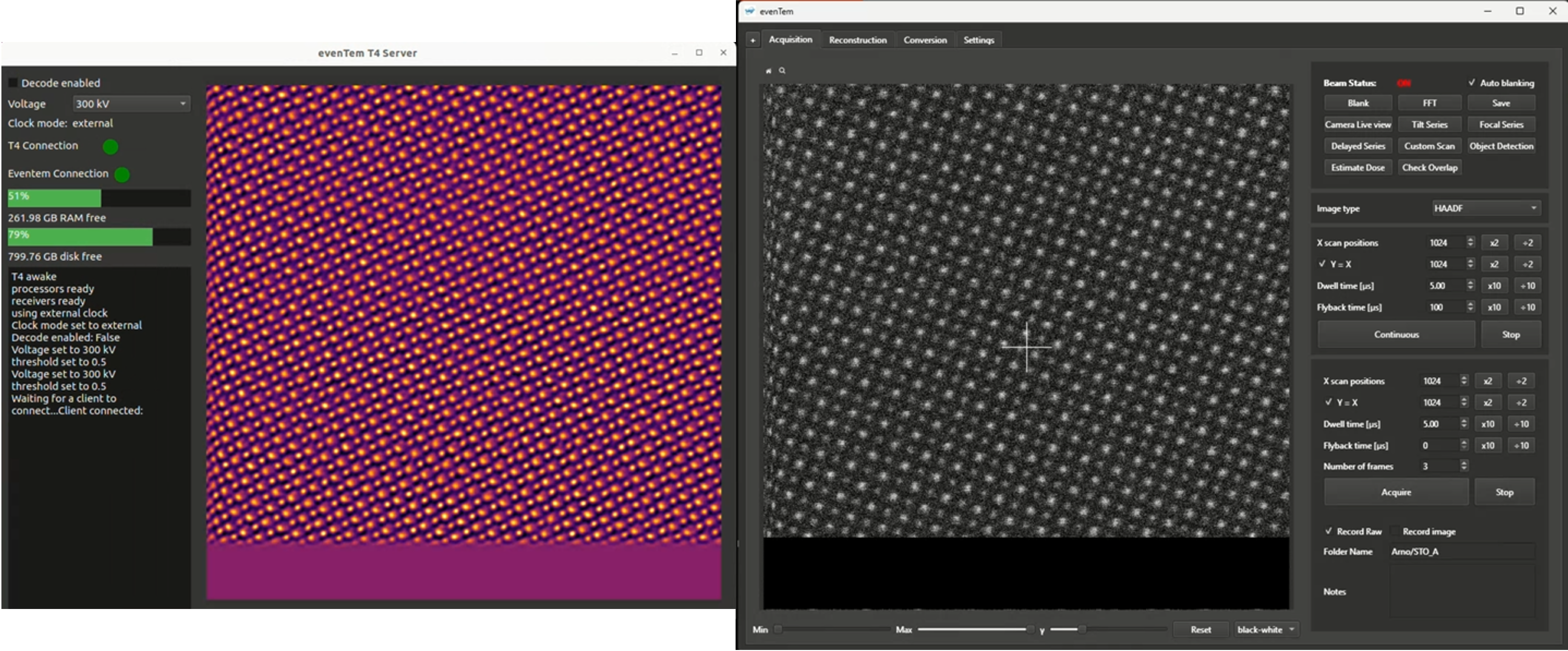}
    \caption{Screenshot of the integration of the real-time 4D-STEM workflow in the experiment operator software (evenTem). The left image panel shows a ptychographic reconstruction using the GPRI framework, reconstructed in real-time from the Timepix4 event stream. The right image panel shows the simultaneous HAADF signal from the physical single-pixel detector.}
    \label{fig:GUI_integration}
\end{figure*}

For the following experimental results sections, the processing is performed using the Merlin T4 control desktop computer. This consists of a 32 core AMD Ryzen Threadripper pro 7975wx processor. The data is written to HDF5 files on a RAM disk with 512 GB of memory. For the ptychographic reconstructions under the GPRI framework\cite{LALANDECROBERT2026114411}, an NVIDIA RTX5080 GPU is used. As noted above, the data processing is performed using the EvenTem software\cite{ANNYS2025}.
\\
\\
Experiments were performed using a Titan 80-300 microscope (Thermo Fisher Scientific), operated at an acceleration voltage of 300~kV and using a probe convergence semi-angle of 21~mrad. A Quantum Detectors scan unit was used to produce a 10~MHz square wave, which was then transformed into a 40~MHz LVDS signal using custom-built electronics, providing the external clock for the Timepix4 detector. 
\\
The combination of few-microsecond dwell times with beam currents in a range of tens of pA places the following measurement well beyond the practical operating regime of conventional frame-based detectors and Timepix3-based event-driven acquisition. It therefore provides a demanding test for the high-throughput event-driven acquisition and processing capabilities enabled by Timepix4.
\\
\\
Videos of the real-time operation at the microscope, showing sustained live processing of the Timepix4 event streams over longer periods, are available as supplementary material and on Zenodo\cite{annys_2026_zenodo}. Figure \ref{fig:GUI_integration} shows a screenshot of such a video, displaying the integration of the workflow into the graphical user interface for experiment control.
\subsection*{Live Annular detectors}
\begin{figure*}[h]
    \centering
    \includegraphics[width=1.0\textwidth]{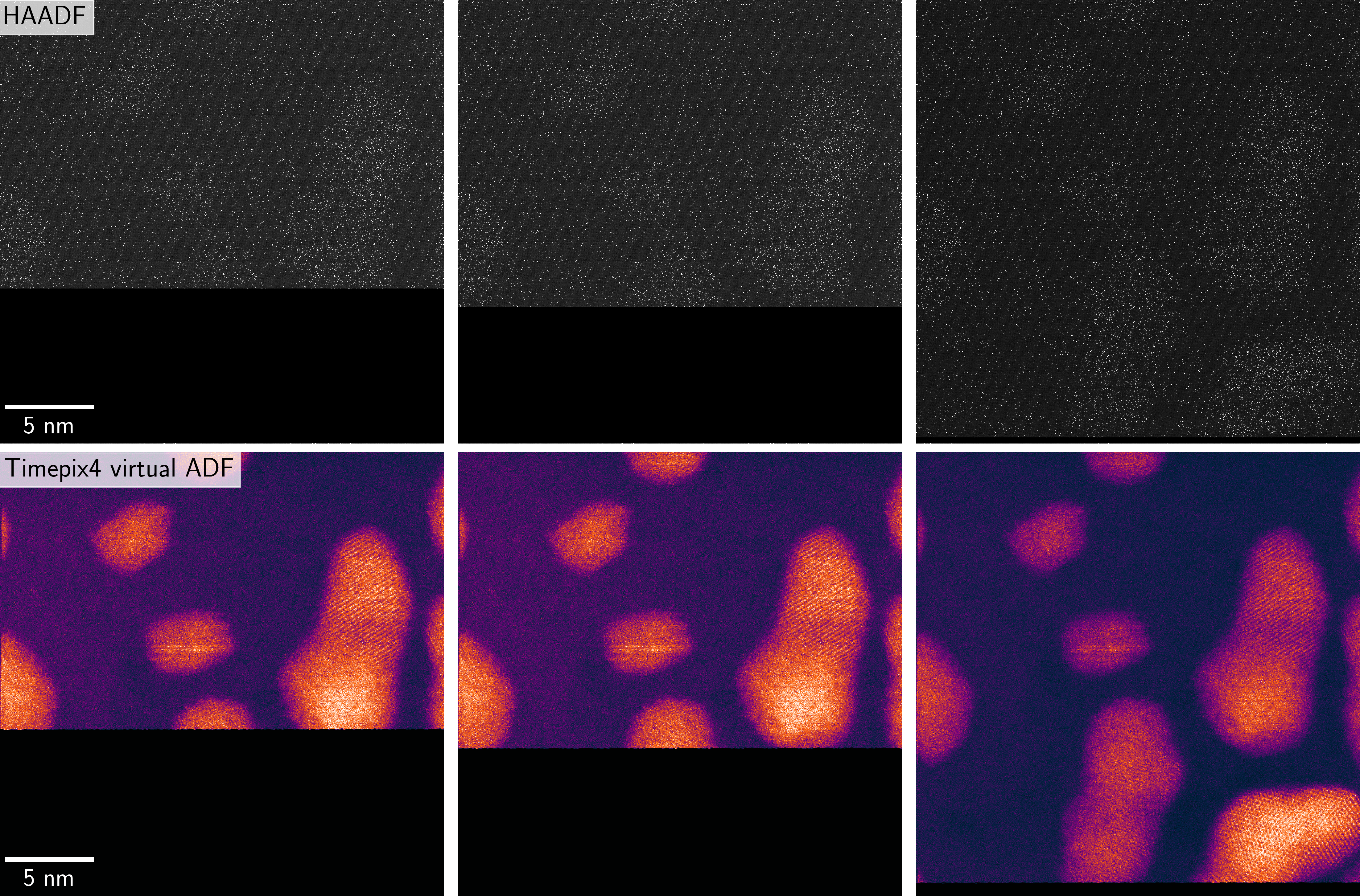}
    \caption{Temporal comparison of signal from the physical HAADF detector with that of the virtual ADF signal from the real-time processing of the Timepix4 event-streams. The scan consisted of 1024 $\times$ 1024 scan points with a pixel dwell time of 5~$\mu$s, resulting in a total scan time of approximately 5.2~s. The top row shows the signal of the physical HAADF signal as the scan progresses. The bottom row shows the simultaneous signal from the real-time 4D-STEM reconstruction. The first column indicates an arbitrary moment in the scan progression, while the second column shows the next update of the 4D-STEM reconstruction, indicating the fast update rate. The final column shows the last render from this scan. There is effectively no latency between the 2 signals, while the reconstructed signal is strikingly more informative. For videos of live operation over sustained periods, the reader is referred to the supplementary material.}
    \label{fig:fig_virtual_stem}
\end{figure*}

Although virtual annular detector reconstructions from 4D-STEM offer limited advantages over conventional hardware annular detectors, they provide a useful benchmark for the processing pipeline, because each detected event contributes only a single addition to the reconstructed image.
Their availability is also relevant because the 4D-STEM detector and hardware annular detectors are usually in different planes in the microscope column, such as e.g. before and after an energy filter. In such configurations, substantially different camera lengths may be required to achieve the desired collection angle range at the two detector planes. As a result, the hardware annular detectors may not provide the desired angular range under the optical conditions used for the 4D-STEM acquisition, thus leaving the operator without a suitable real-time imaging signal. Virtual annular detectors provide a flexible alternative by allowing the collection range and shape to be selected flexibly.
\\
For the experimental demonstration, a standard gold nanoparticle calibration grid was used. Continuous $1024 \times 1024$ scans were performed with a dwell time of 5~$\mu$s per probe position. The beam current was estimated to be 26~pA, with a mean cluster size of 2 pixels per electron. This beam current was not limited by the detector or processing workflow.
Figure \ref{fig:fig_virtual_stem} shows the signal provided by a real-time 4D-STEM annular dark-field (ADF) reconstruction, alongside the one generated by the physical HAADF detector, at the nearest rendering moment. The first column indicates an arbitrary moment in the scan progression, while the second column shows the next update of the 4D-STEM reconstruction, highlighting the high update rate. The final column shows the last render from this scan, before the next one is visualized. There is no significant delay between the 2 signals. In fact, the reconstruction from the 4D-STEM signal often runs ahead of the signal from the physical detector, justifying the claim of zero latency for the proposed 4D-STEM workflow. The camera length was set up to have a desirable illumination of the pixelated detector, where the diameter of the bright field disk is approximately half the detector size. As a consequence of those experimental choices, the signal from the physical HAADF detector shows severe noise corruption stemming from both Poisson noise due to an ineffective collection range, and from additional noise sources such as electronic noise. The virtual ADF signal is significantly more informative, showing clear atomic resolution, due to the flexibility in collection ranges and the efficiency of detection.
\subsection*{Live RiCOM}
\begin{figure*}[h]
    \centering
    \includegraphics[width=1.0\textwidth]{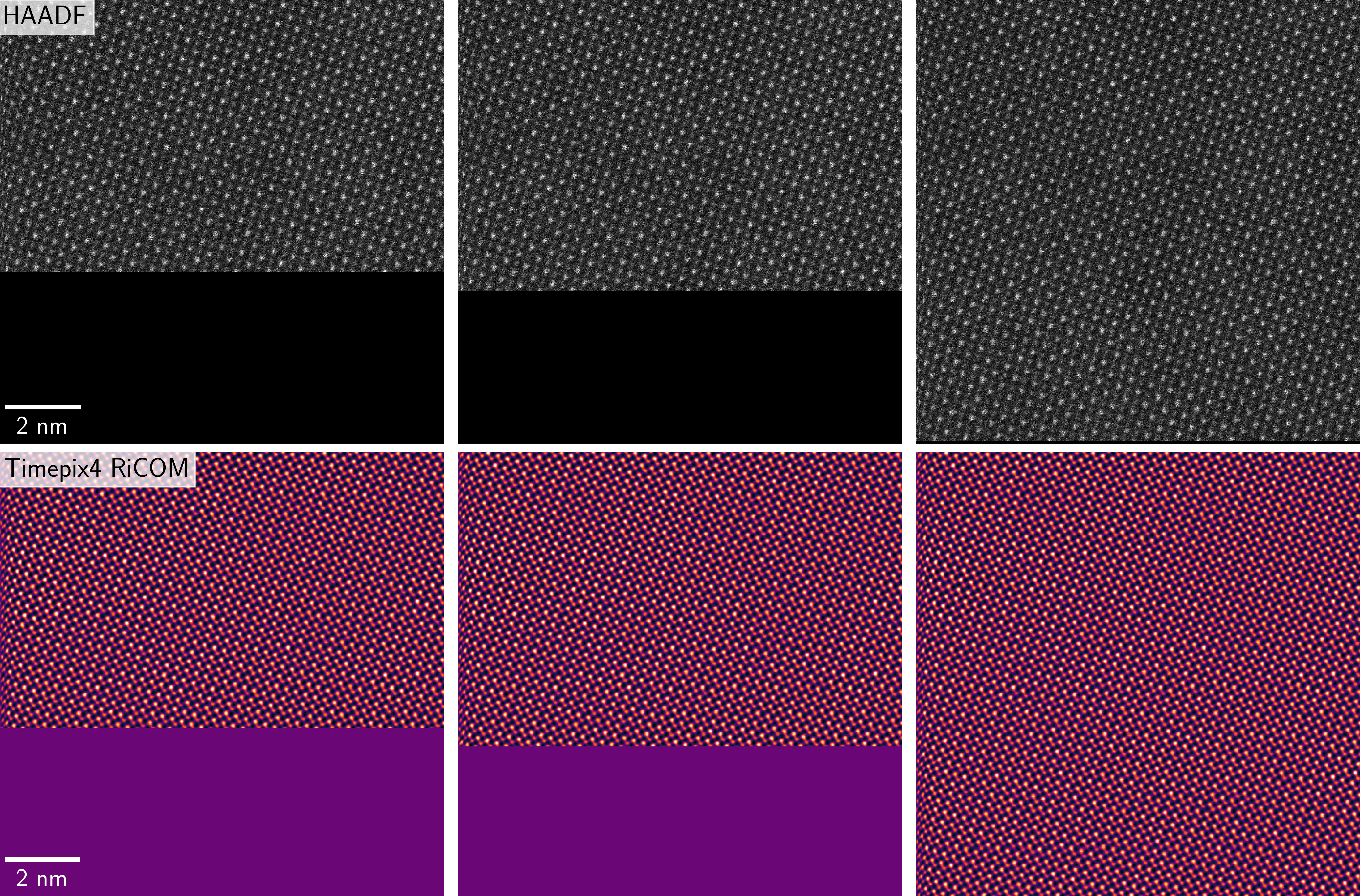}
    \caption{Comparison of signal from the physical HAADF detector with that of real-time integrated center of mass processing of the Timepix4 event-streams. The scan consisted of 1024 $\times$ 1024 scan points with a pixel dwell time of 5~$\mu$s, resulting in a total scan time of approximately 5.2~s. The top row shows the signal of the physical HAADF signal as the scan progresses. The bottom row shows the simultaneous signal from the real-time 4D-STEM reconstruction. The first column indicates an arbitrary moment in the scan progression, while the second column shows the next update of the 4D-STEM reconstruction, indicating the fast update rate. The final column shows the last render from this scan. There is effectively no latency between the 2 signals, while the reconstructed signal provides an increased signal-to-noise ratio and improved contrast, especially for the light element columns. For videos of live operation over sustained periods, the reader is referred to the supplementary material.}
    \label{fig:fig_ricom}
\end{figure*}

RiCOM \cite{Yu2022a} provides a particularly efficient approach to real-time 4D-STEM imaging, while retaining many of the advantages associated with more advanced ptychographic methods such as SBI and WDD, including their dose-efficiency and sensitivity to light elements\cite{Kavak2025,chennit2026pre}. The computational cost of the reconstruction is exceptionally low: each detected event requires only three additions in updating the center of mass (COM) maps. The RiCOM image, which provides a representation of the projected electrostatic potential, can then be obtained from those COM maps through a local integration, performed in parallel with the treatment of events. This combination of low per-event computational cost and direct event-driven processing makes RiCOM particularly well suited to interactive 4D-STEM. Moreover, its computational simplicity allows real-time operation without requiring a GPU, in contrast to the slightly more computationally demanding alternative for phase contrast, represented by GPRI.
\\
For the experimental demonstration, a strontium titanate thin film\cite{Pryds_2026} is used. Like in the previous case, repeated $1024 \times 1024$ scans were performed with a dwell time of 5~$\mu$s per probe position. The beam current was estimated to be 37~pA. Again, this beam current was not limited by the detector or processing workflow. Figure \ref{fig:fig_ricom} shows the result of the RiCOM reconstruction, again next to the signal obtained from the physical HAADF detector, at the nearest rendering moment. The figure follows the same format as figure \ref{fig:fig_virtual_stem}. There is no significant delay between the 2 signals, demonstrating that the low latency of the workflow can be preserved in more advanced imaging methods. Moreover, the RiCOM image clearly highlights the oxygen columns in this projected atomic structure, which are not visible in the HAADF image due to the poor light element sensitivity.
\subsection*{Live ptychography under the GPRI framework}
\begin{figure*}[h]
    \centering
    \includegraphics[width=1.0\textwidth]{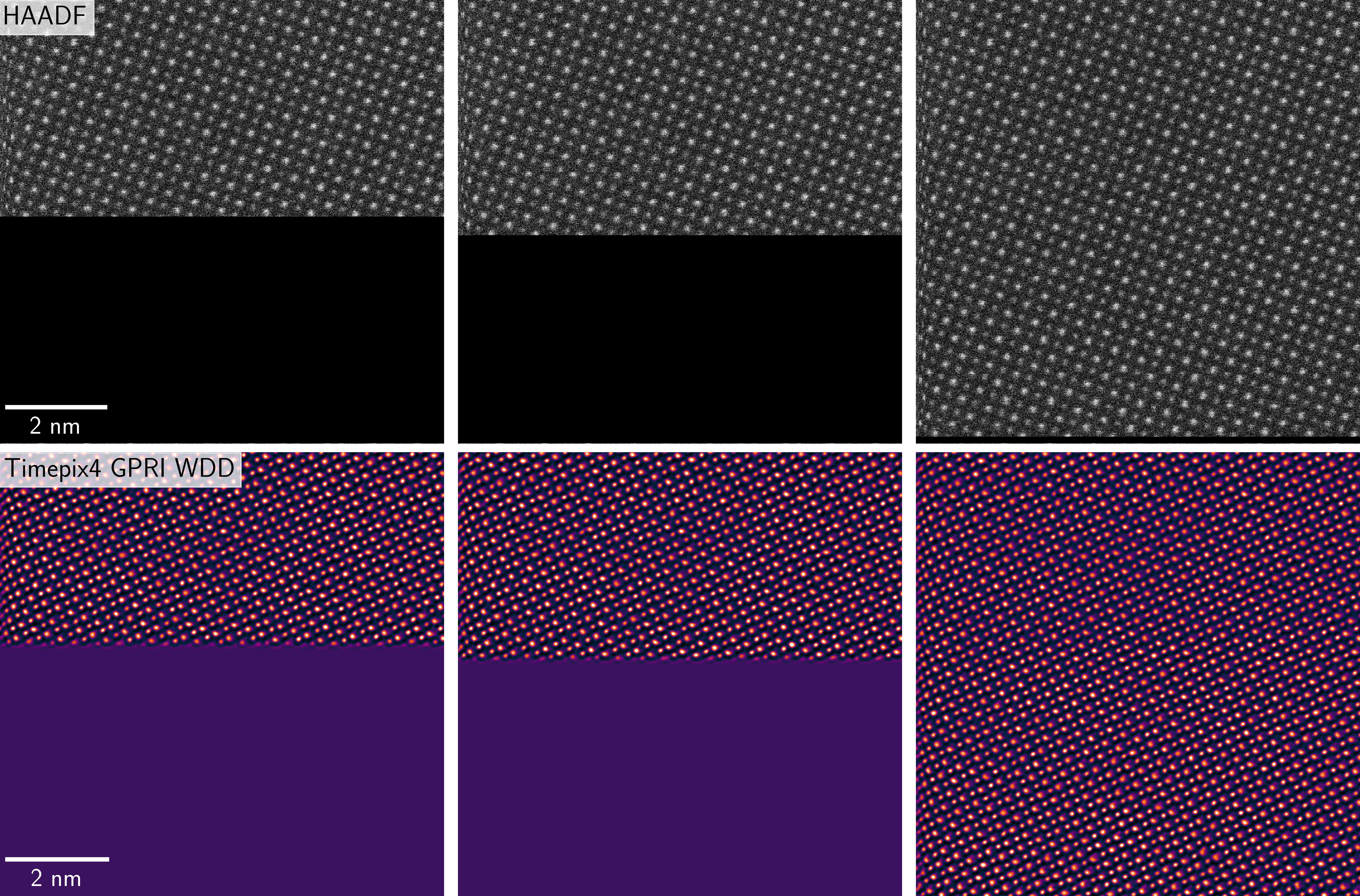}
    \caption{Temporal comparison of signal from the physical HAADF detector with that of GPRI WDD real-time processing of the Timepix4 event-streams. The scan consisted of 1024 $\times$ 1024 scan points with a pixel dwell time of 5~$\mu$s, resulting in a total scan time of approximately 5.2~s. The top row shows the signal of the physical HAADF signal as the scan progresses. The bottom row shows the simultaneous signal from the real-time ptychographic reconstruction. The first column indicates an arbitrary moment in the scan progression, while the second column shows the next update of the 4D-STEM reconstruction, indicating the fast update rate. The final column shows the last render from this scan. There is very little latency between the 2 signals, while the reconstructed signal provides an increased signal-to-noise ratio and improved contrast, especially for the light element columns. For videos of live operation over sustained periods, the reader is referred to the supplementary material.}
    \label{fig:fig_GPRI}
\end{figure*}

The manner in which an analytical ptychography process can be conducted event by event has been described in detail in a prior publication \cite{LALANDECROBERT2026114411}. In summary, spatially limited kernels containing the pre-calculated contribution of electron events are summed in a continuously updated reconstruction result, involving a region of N $\times$ N pixels around the scan position associated with the event. Since this reduces the reconstruction to a summation of independent contributions, it can be done in parallel, in a thread-safe way. An analysis of the computational cost of such a process, in terms of the number of floating-point operations per second (FLOPS), has already been reported in ref. \cite{ANNYS2025}. In summary, it scales only with the kernel size and the event rate of the detector, and might range from GFLOPS to 100 TFLOPS in realistic conditions, which is within the performance range of modern GPUs. Reconstruction rates that exceed the acquisition rates of datasets recorded with a Timepix3 chip at beam currents of approximately 1 pA have been demonstrated in ref. \cite{LALANDECROBERT2026114411}. Real-time operation at the microscope using a higher beam current had not been demonstrated yet.
\\
For the experimental demonstration, again, continuous $1024 \times 1024$ scans were performed with a dwell time of 5~$\mu$s per probe position, imaging the same SrTiO$_{3}$ specimen as in the RiCOM example above. The beam current was estimated to be 11~pA. Kernels of size 32 $\times$ 32 pixels were used, requiring 135~GFLOPS at a beam current of 11~pA, meaning higher beam currents are certainly possible. Figure \ref{fig:fig_GPRI} shows the micrograph resulting from GPRI-based direct ptychography, next to the signal of the physical HAADF detector at the nearest rendering moment. Specifically, the GPRI signal in figure \ref{fig:fig_GPRI} depicts the quantity $Im\left[\,\gamma\,T^{WDD}\left(\vec{r}\right)\,\right]$, where $\gamma$ is a scale factor and $T^{WDD}\left(\vec{r}\right)$ is a specimen transmission function, as described in detail in \cite{LALANDECROBERT2026114411}. The contrast of this image approximately relates to the phase shift. The figure follows the same format as figures \ref{fig:fig_virtual_stem} and \ref{fig:fig_ricom}. There is no significant delay between the 2 signals, demonstrating that the remarkably low latency of the workflow can be preserved even for this form of electron ptychography. 
\\
Similar to the RiCOM result in figure \ref{fig:fig_ricom}, the WDD reconstruction in figure \ref{fig:fig_GPRI} shows improved contrast, signal-to-ratio and light element sensitivity compared to the HAADF signal. Although possible, no computational correction of aberrations was attempted here. A significant benefit of performing the ptychography with a focused probe in a dense scan is that a useful HAADF image, with its well-understood and interpretable contrast is retained.
\section*{Discussion}
The work presented in this publication demonstrates the experimental realization of large 4D-STEM scans with microsecond dwell times, while maintaining standard beam currents, combined with practically zero-latency data processing, in a sustained real-time operation. Hence, the existing user experience of flexible operation with a single pixel detector is combined with the powerful capabilities of 4D-STEM imaging and electron ptychography. As a result, the information available during experiments can transition from ineffective analog signals, such as HAADF-STEM, to dose-efficient, high contrast, images with light elements sensitivity. The availability of the desired contrast mechanism, such as e.g. phase contrast, also allows the operator to account for the differences in optimal focus between different methods. The live availability of those valuable measurement results can make the difference between a nearly blind operation of the electron microscope to effective and timely decision-making. These developments make it possible to introduce a new, highly advantageous, experimental workflow in future STEM applications, holding great promise for incorporation in increasingly automated and information-driven workflows\cite{Kalinin2023,CPYu2026}.
Finally, another important benefit of the event-driven methodology reported here is that it does not require hardware beyond standard lab infrastructure, being based on a single desktop computer. This avoids the need of incorporating high-performance computing (HPC) infrastructure into the 4D-STEM workflow\cite{Mukherjee2022,Welborn2024}. This is a very significant consideration regarding the possibility of widespread adaptation and accessibility, which is heavily determined by factors such as cost and barrier to entry. Additionally, the fact that the proposed workflow retains the raw data used for the live measurement allows maximal flexibility in deciding post-hoc to store the raw data and further process it with other methods, possibly using HPC.
\section*{Conclusion}
We have developed a high-throughput interactive 4D-STEM workflow that exploits the parallel data output of the Quantum Detectors Merlin T4 detector, based on the Timepix4 chip. By providing a common clock between the scan unit and detector, the probe position can be directly determined from the time-of-arrival of each event, eliminating the need for trigger-based synchronization or the sorting of events between parallel data streams. This allows the 32 detector data streams to be processed independently and in parallel, thus providing a scalable architecture for real-time event-driven 4D-STEM.

Using this pipeline, we demonstrated real-time processing of event-driven 4D-STEM data, including virtual annular detectors and RiCOM, at microsecond dwell times and with standard beam currents exceeding tens of pA. These operating conditions extend substantially beyond those demonstrated with Timepix3-based workflows, where the available event throughput limited operation to currents of a few pA only, and beyond the operating regime of a frame-based acquisition as well. We furthermore demonstrate real-time direct electron ptychography using WDD, under the GPRI framework, using megapixel scans, microsecond dwell times and double-digit beam currents.

Together, these results demonstrate that the increased throughput of Timepix4 detectors can be combined with a suitably designed acquisition and processing architecture to provide live 4D-STEM at acquisition conditions previously inaccessible. In particular, the ability to perform direct ptychographic reconstruction, with sufficiently low latency to provide immediate experimental feedback, brings phase contrast imaging closer to the interactive experience traditionally associated with conventional STEM signals. More broadly, the presented architecture provides a framework for making high-throughput event-driven 4D-STEM a practical and interactive mode of operation that holds promise to become the driving tool of future experiments.

\section*{CRediT authorship contribution statement}

        A.A.: conceptualization, implementation, experiments, manuscript writing.  H.L.L.R.: conceptualization, implementation, manuscript revision. J.V.: conceptualization, supervision, manuscript revision.

\section*{Acknowledgments}

    The authors thank Giulio Crevatin and Matus Krajnak from Quantum Detectors for their technical support.
    Prof. Nini Pryds (Technical University of Denmark) and his group are gratefully acknowledged for providing the SrTiO$_{\text{3}}$ specimen.
    \\
    A.A., H.L.L.R and J.V. acknowledge funding from the European Union under grant agreement ID 101094299 (IMPRESS). Views and opinions expressed are however those of the authors only and do not necessarily reflect those of the European Union or the European Research Executive Agency (REA). Neither the European Union nor the granting authority can be held responsible for them.
    \\
    H.L.L.R. acknowledges additional funding from the Research Foundation- Flanders (FWO, Belgium), under granted postdoctoral fellowship No. 12A9V27N.
    
\section*{Declaration of conflicts of interest}
    The authors declare no conflicts of interest.

\section*{Supplementary videos}
Videos of the real-time operation at the microscope, showing sustained live processing of the Timepix4 event streams over longer periods, are available as supplementary material and on Zenodo\cite{annys_2026_zenodo}.

\bibliography{biblio,morebib}

\end{document}